# Direct and indirect electrocaloric measurements using multilayer capacitors

S. Kar-Narayan and N. D. Mathur

Department of Materials Science, University of Cambridge, Cambridge, CB2 3QZ, UK

*Abstract*

We report the discovery of serendipitous electrocaloric effects in commercial multilayer capacitors based on ferroelectric $BaTiO_3$. Direct thermometry records ~0.5 K changes due to 300 kV cm$^{-1}$, over a wide range of temperatures near and above room temperature. Similar results are obtained indirectly, via thermodynamic analysis of ferroelectric hysteresis loops. We compare and contrast these two results. Optimised electrocaloric multilayer capacitors could find applications in future cooling technologies.

## 1. Introduction

Mathur and coworkers recently showed that industrially manufactured multilayer capacitors (MLCs) based on piezoelectric $BaTiO_3$ inadvertently show strain-mediated magnetoelectric effects[1-3], because the interdigitated electrodes are now based on magnetostrictive Ni to replace more expensive non-magnetic Ag-Pd alloys. Here we show that similar MLCs display electrocaloric (EC) effects even though they were not designed for this purpose either. The MLC geometry is attractive as the electrodes provide both thermal and electrical access to a significant quantity of EC material with no substrate.

EC effects are electrically driven changes in temperature $\Delta T$, and they are largest near phase transitions. Small effects (e.g. $\Delta T \sim 2$ K) arise in bulk ceramics[4-13], but larger effects (e.g. $\Delta T = 12$ K) in ceramic[14-16] or polymer[17] thin films may be possible as thin films can support larger electric fields. EC effects (e.g. $\Delta T \sim 1$ K)[7] in thick films have also been demonstrated using an interdigitated multilayer geometry[7,9], but only two reports exist and multilayer fabrication is challenging. Here we demonstrate EC effects in mass-produced MLCs that are readily available for one cent or less. We show that indirect thermodynamic analysis of ferroelectric hysteresis loops yields same strength EC effects as direct measurements.

## 2. Experimental details

Our MLC (Fig. 1a) is a commercially available electronic component (1210ZG226ZAT2A, AVX, Northern Ireland), where the dielectric is Y5V ($BaTiO_3$ doped for high permittivity and reasonable temperature stability), and the electrodes are based on nickel. The 10 V maximum voltage (i.e. ~15 kV cm$^{-1}$), recommended by the manufacturers in order to avoid poling and hysteresis which would compromise capacitor function in standard electronic applications, is well below the breakdown field and we were able to reliably[2] apply up to 300 kV cm$^{-1}$ (where dielectric loss $\leq$5% at 120 Hz)[18]. Indirect and direct EC measurements were performed using apparatus (Fig. 1b) held at ~$10^{-3}$-$10^{-2}$ Torr in a small vacuum tube, with one flange for electrical feedthroughs and the other for a rotary pump. A copper block with a recess, covered everywhere with

Kapton tape for electrical insulation, was wound with a constantan-wire heater. The temperature could be controlled to 0.1 K using feedback from a Pt-100 thin-film thermometer (not visible in Fig. 1b). The MLC was mounted in the recess on one of its terminals, via a weak thermal link comprising a 1 mm-thick wooden spacer coated with GE varnish for mechanical adhesion. Two Pt-100 thin-film thermometers (Fig. 1b) were mounted using N-grease for good thermal contact, one on the exposed MLC terminal (sample thermometer), the other on the recess wall (reference thermometer). All electrical connections were made using thin copper wire (diameter ~ 0.12 mm). Similar results to those we present here were obtained for nominally identical and other MLCs.

## 3. Results

The MLC was characterized at and above room temperature by recording electrical displacement $D$ versus electric field $E$ via the constant-current method[19,20], and using this data to predict EC temperature change. The ferroelectric hysteresis loops (e.g. Fig. 2a) are slimmer than those we observed for several other commercially available MLCs, justifying MLC selection here. Leakage currents (Fig. 2b) at high field (300 kV cm$^{-1}$) reach ~1 µA at ~360 K, corresponding to a small ~1.5% correction to $D$, and for what follows we reject data above this temperature. Below ~360 K, $D(E)$ loops (upper branches, $E > 0$) were used to establish $D(T)$ at selected $E$ (Fig. 2c). The indirect thermodynamic method[6] was then used to predict EC temperature change $\Delta T^I$, where:

$$\Delta T^I = -\frac{1}{\rho}\int_{E_1}^{E_2}\frac{T}{c}\left(\frac{\partial D}{\partial T}\right)dE, \tag{1}$$

for changes in $E$ from $E_1$ to $E_2$. We assume the Maxwell relation $(\partial D/\partial T)_E = (\partial S/\partial E)_T$, $c$ is specific heat capacity and $\rho$ is density. For $\Delta E = E_2 - E_1 = -300$ kV cm$^{-1}$ ($E_2 = 0$), we find $-\Delta T^I_{Y5V} \sim 1$ K (dotted line, Fig. 2d) in the Y5V dielectric of the MLC, assuming[21] $\rho_{Y5V} = 5840$ kg m$^{-3}$ and a constant $c_{Y5V} = 434$ J K$^{-1}$ kg$^{-1}$ for Y5V ceramics ($c_{Y5V}$ varies by ~5% in 300 K $< T <$ 360 K)[22]. By taking into account the thermal mass of the nickel-based electrodes, $|\Delta T^I_{Y5V}|$ is reduced to yield a predicted MLC temperature change of magnitude $|\Delta T^I_{MLC}|$, where:

$$\Delta T^I_{MLC} = \left[1 - \left(\frac{c_e \rho_e V_e}{c_e \rho_e V_e + c_{Y5V}\rho_{Y5V}V_{Y5V}}\right)\right]\Delta T^I_{Y5V} = 0.68 \cdot \Delta T^I_{Y5V}. \qquad (2)$$

The MLC in effect comprises $n = 200$ parallel-plate capacitors of effective length $L \sim 3.3$ mm and effective width $w \sim 2.56$ mm. Each Y5V layer has thickness $d_{Y5V} \sim 6.5$ μm, and so the total dielectric volume $V_{Y5V} = n v_{Y5V} = nLw d_{Y5V} = 10.9$ mm$^3$. Each electrode layer has thickness $d_e \sim 2.0$ μm, and so the total electrode volume $V_e = (n+1)v_e = (n+1)Lwd_e = 3.4$ mm$^3$. We assume that both the electrode specific heat capacity $c_e = 429$ J K$^{-1}$ kg$^{-1}$ and electrode density $\rho_e = 8907$ kg m$^{-3}$ take the values[23] for pure nickel. The maximum EC cooling of the MLC ($-\Delta T^I_{MLC}$) is thus ~0.9 K (solid line, Fig. 2d). Ferroelectric hysteresis losses[14] could reduce this figure by up to ~0.06 K. Joule heating would only weakly affect direct measurements of the EC temperature changes predicted here, as the maximum leakage (~1 μA, Fig. 2b) at the maximum temperature of interest (~360 K, Figs. 2b-d) corresponds to a heating of just ~0.009 K over the relevant time step of 3 s (see Fig. 3a, later). Note that using the lower branches of hysteresis loops (Fig. 2a) would alter the predicted EC temperature changes (Fig. 2d) by $\leq \pm 10\%$ down to 325 K, and increase them to more optimistic values by up to 60% at lower temperatures.

Directly measured temperature changes in the MLC are presented in Fig. 3. At room temperature (Fig. 3a), the MLC thermometer records a sharp increase over ~ 3 s due to the application of 300 kV cm$^{-1}$. Most of the heat subsequently leaks away on a timescale (~100 s) that depends primarily on the weak thermal link between the MLC and the copper block via the wooden spacer. The subsequent removal of the electric field produces a sharp decrease in temperature over ~ 3 s, after which heat leaks back into the MLC primarily from the copper block on the same ~100 s timescale. The sharp changes primarily represent direct measurements of EC temperature change $\Delta T_{\text{MLC}}^{\text{D}}$, and at various starting temperatures we find $\left| \Delta T_{\text{MLC}}^{\text{D}} \right|$ ~ 0.5 K (Fig. 3b). The magnitude of the sharp heating step exceeds the magnitude of the sharp cooling step by no more than ~0.02 K (Fig. 3b), which we attribute to small losses associated with e.g. hysteresis and Joule heating. Note that although the MLC thermometer does not quite return to the starting temperature when the field is on, this apparent discrepancy is very small, and is not well resolved in these experiments.

## 4. Discussion

Values of $|\Delta T_{\text{MLC}}^{\text{D}}|$ and $|\Delta T_{\text{MLC}}^{\text{I}}|$ agree to within much better than an order of magnitude at each measurement temperature, as expected[6,12,13]. The slight discrepancies could be due to the approximations associated with the indirect method, or the influence on the direct measurements of unaccounted thermal mass (unaddressed dielectric, Pt-100 thermometer, N-grease, wiring, etc.). The small peaks in $|\Delta T_{\text{MLC}}^{\text{D}}|$ and $|\Delta T_{\text{MLC}}^{\text{I}}|$ cannot be reconciled with the broad 290 K peak in the dielectric constant[18], or our own heat capacity data where no peaks are discernible, but this is an issue of detail, and our purpose here is to show that the MLCs display EC effects.

The magnitude of our EC temperature changes $|\Delta T_{\text{MLC}}^{\text{D}}| \sim 0.5\,\text{K}$ are low with respect to ~12 K predictions[14,17] for thin-film PZT and P(VDF-TrFE). This is because good EC performance is not expected in our Y5V dielectric for three reasons. First, films of the parent $BaTiO_3$ compound are predicted by the indirect method to show only small EC temperature changes (~1 K for $\Delta E = 100\,\text{kV cm}^{-1}$)[24]. Second, we could not access the large driving fields ($\leq 1000\text{-}2000\,\text{kV cm}^{-1}$)[14,17] required for large EC effects as our MLC is based on thick films. Third, doping suppresses phase transitions, which is good for standard MLC applications, but bad for EC effects. Also, the active regions of the MLC are clamped by the surrounding unaddressed dielectric, which further suppresses phase transitions and therefore EC performance[14,24,25].

EC performance would be significantly improved by better materials selection[26], suggesting the possibility[26] of employing MLCs in EC refrigerators[27] where cooling is achieved via both MLC terminals. The previous work[7,9] using multilayers to determine EC materials performance did not regard the multilayers as devices for applications: EC temperature changes for the multilayers themselves were not reported, and the ability of the electrodes to conduct heat was not identified consistent with mounting the thermometer away from the terminal, unlike here.

**5. Conclusions**

We have shown that commercially available MLCs based on ferroelectrics serendipitously show EC effects. The MLC geometry is ideal for direct EC measurements and applications because the interdigitated electrodes provide both electrical and thermal access to a macroscopic quantity of EC material that can be organised in thin-film form where large EC effects are expected[14-17]. Materials and geometric optimization would enhance the potential for small-scale (e.g. on-chip) cooling applications. Higher cooling powers for refrigeration and air-conditioning could be achieved using a large number MLCs.

**Acknowledgements**

We thank Casey Israel for aiding MLC selection, and we thank Suman-Lata Sahonta for assistance with MLC cutting and polishing. This work was funded by UK EPSRC grant EP/E03389X.

Corresponding author (NDM): ndm12@cam.ac.uk

**Figures**

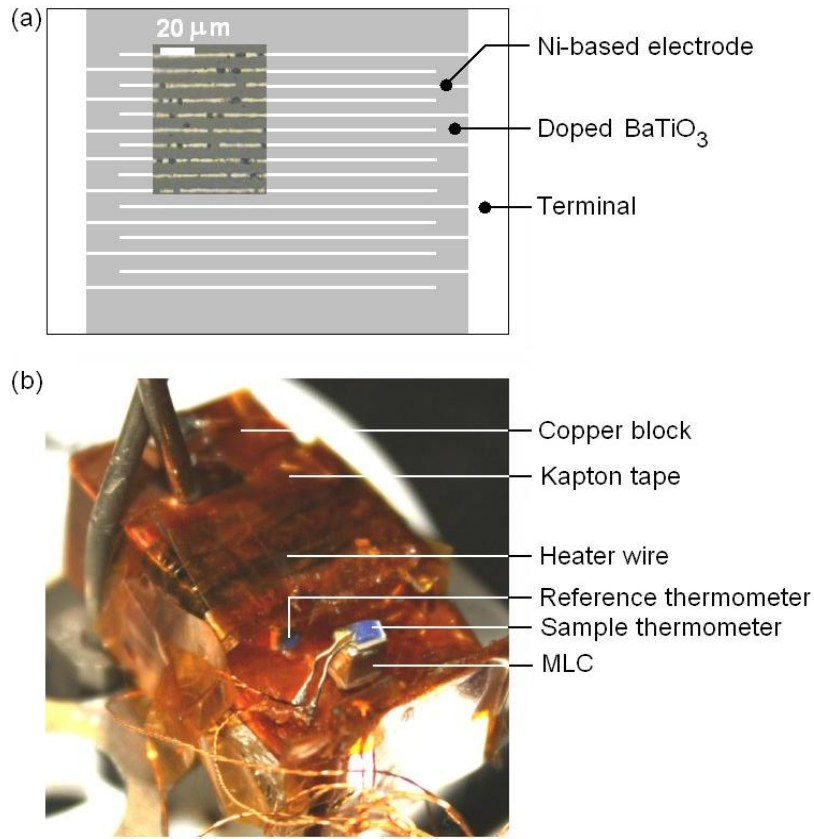

Fig. 1. MLC and experimental apparatus. (a) MLC cross-sectional schematic with overlay obtained by optical microscopy after sawing and polishing. In reality, there are 200 dielectric layers of thickness $d_{Y5V}$ ~ 6.5 μm, separated by Ni-based electrodes of thickness $d_e$ ~ 2.0 μm. The active area of each dielectric layer is ~8.5 mm$^2$. (b) Experimental set-up, comprising MLC and sample thermometer, mounted on a bespoke variable-temperature copper block with a reference thermometer. Experiments were performed at ~$10^{-3}$-$10^{-2}$ Torr in a small vacuum chamber.

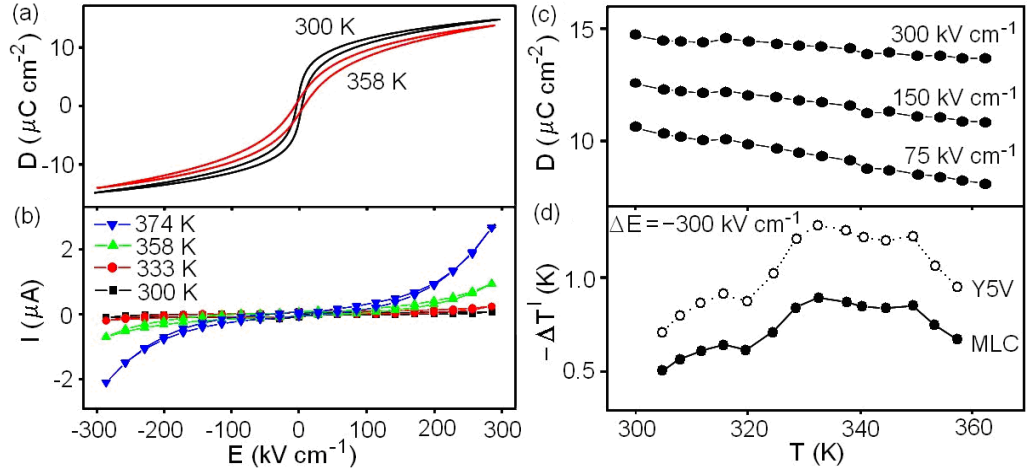

Fig. 2. MLC characterization via indirect EC measurements. (a) Electric displacement ($D$) versus electric field ($E$) at selected temperatures, taken using the constant-current method[19,20] (±10 μA, period ~100 s). (b) Steady-state leakage current $I$ for these and selected other loops (settle time ~300 s). (c) $D(T)$ at selected $E$, extracted from $D(E)$ loops (upper branches, $E > 0$) at 16 measurement temperatures. Using Equation 1, data of this type were used to predict (d) EC temperature changes $-\Delta T^I_{Y5V}$ (dotted line) in the Y5V dielectric for $\Delta E = E_2 - E_1 = -300$ kV cm$^{-1}$ ($E_2 = 0$). The magnitude of $-\Delta T^I_{Y5V}$ is larger than the magnitude of the predicted MLC temperature change $-\Delta T^I_{MLC}$ (solid line) because of electrode thermal mass (Equation 2). Data in (d) smoothed by five-point averaging.

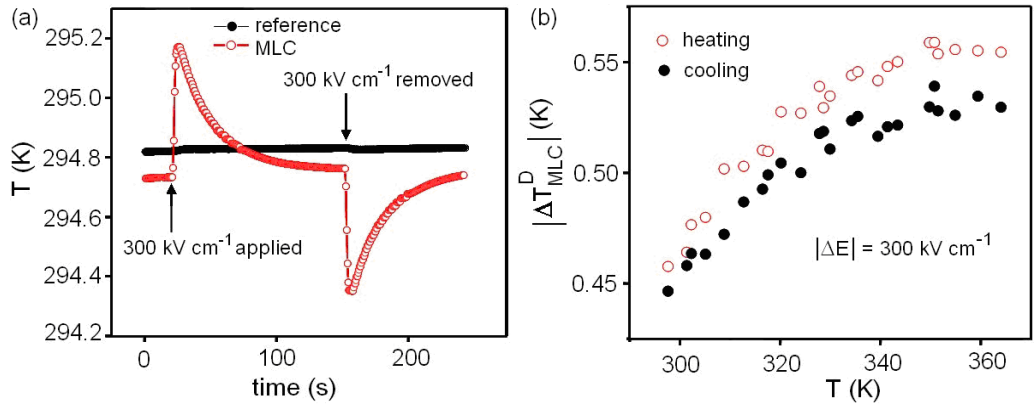

Fig. 3. Direct EC measurements of the MLC. (a) At room temperature, the MLC thermometer records temperature changes over time due to the application and subsequent removal of 300 kV cm$^{-1}$. These changes are not quite seen in the reference thermometer. The sharp heating and cooling jumps are primarily due to the EC effect. Small losses associated with e.g. hysteresis and Joule heating render the heating jumps slightly larger than the cooling jumps, as seen in (b) where we plot jump magnitude $\left|\Delta T_{\mathrm{MLC}}^{\mathrm{D}}\right|$ at various MLC starting temperatures.